# Cognitive network science for understanding online social cognitions: A brief review

*Massimo Stella, Department of Computer Science, University of Exeter, EX4 4QF Exeter, UK*

*Social media are digitalising massive amounts of users' cognitions in terms of timelines and emotional content. Such Big Data opens unprecedented opportunities for investigating cognitive phenomena like perception, personality and information diffusion but requires suitable interpretable frameworks. Since social media data come from users' minds, worthy candidates for this challenge are cognitive networks, models of cognition giving structure to mental conceptual associations. This work outlines how cognitive network science can open new, quantitative ways for understanding cognition through online media, like: (i) reconstructing how users semantically and emotionally frame events with contextual knowledge unavailable to machine learning, (ii) investigating conceptual salience/prominence through knowledge structure in social discourse; (iii) studying users' personality traits like openness-to-experience, curiosity, and creativity through language in posts; (iv) bridging cognitive/emotional content and social dynamics via multilayer networks comparing the mindsets of influencers and followers. These advancements combine cognitive-, network- and computer science to understand cognitive mechanisms in both digital and real-world settings but come with limitations concerning representativeness, individual variability and data integration. Such aspects are discussed along the ethical implications of manipulating socio-cognitive data. In the future, reading cognitions through networks and social media can expose cognitive biases amplified by online platforms and relevantly inform policy making, education and markets about massive, complex cognitive trends.*

## INTRODUCTION

A key component of cognition lies in the ability to express ideas through language. Over the centuries, concepts and emotions were retrieved from the human mind, encapsulated in words and then diffused through written and oral media. Only in the last few decades this process accelerated drastically. Online social media, mimicking friendship circles, revolutionised people's ways to speak their minds, structuring their stances, knowledge and perceptions through social discourse, timelines and posts (Stella et al. 2018; Gallagher et al. 2019; Hills 2019; Firth et al. 2019). Social media have a strong cognitive component because they are mainly made of knowledge, i.e. emotions and ideas, flowing from user to user along social ties (Dodds et al. 2011, Hills 2019, Corallo et al. 2020). This information flow is used by online users to either build or express their own experience in ways that are not fully known yet (cf. Hills 2019).

The growth of social media usage is creating datasets increasingly larger, more complex and more complicated to integrate (cf. Cresci et al. 2017, Stella et al. 2018, Cinelli et al. 2020, Botta et al. 2020) but at the same time also way more informative about language and its cognitive reflection within the human mind (Vitevitch 2019, Stella 2020). Investigating social media data means achieving a better understanding of cognitive mechanisms related to information processing, seeking and contagion by checking the content and choices produced by millions of online users every day (Firth et al. 2019, Hills 2019).

Social media data comes as a great opportunity but also as an open challenge. It requires modelling frameworks capable of highlighting the structure of knowledge flowing on social ties. Recently the science of complex networks stepped up from purely social analyses and started encompassing the cognitive dimensions of social media (Del Vicario et al. 2016, Gallagher et al. 2018, Lydon-Staley et al. 2019, Stella et al. 2020, Stella 2020, Radicioni et al. 2020). Whereas the first Big Data explorations of social media successfully investigated millions and millions of user interactions (cf. Ferrara et al. 2016, Cresci et al. 2017, Botta et al. 2020), they did not unveil the cognitive content of such exchanges, i.e. if they included specific emotions or ideas or rather hate speech or stereotypical perceptions (Hills 2019). Unearthing the cognitive dimension of social media knowledge can inform on a variety of features that are not captured by social interactions only, like:

(i) identifying the stance of users discussing a given event, person or idea, e.g. the gender gap or an exhibition (Mohammad et al. 2016, Corallo et al. 2020, Stella 2020);

(ii) profiling what type of emotions were vehiculated by massively resharings, e.g. after the lockdown release Italian Twitter users re-tweeted more messages eliciting stronger patterns of fear (Stella 2020b);

(iii) reconstructing the flow of emotions and stances over time, providing structure to the narrative exchanged by users over time, e.g. identifying how users revealed their own experiences in the #MeToo movement (Gallagher et al. 2020);

(iv) assessing in a quantitative way how users framed specific news, with relevance for understanding how social bots (Ferrara et al. 2016, Stella et al. 2018), fake accounts (Cresci et al. 2017) and trolls (Monakhov 2020) influenced other users;

(v) profile also online users in terms of their digital footprint based on the language they used online, e.g. assess personality traits and information seeking patterns (Hills 2019, Lydon-Staley et al. 2020).

This complex landscape must be investigated by giving structure to knowledge in social discourse, a task achievable by reconstructing conceptual associations between ideas. Cognitive network science (Siew et al. 2019) represents an ideal candidate for this. At the fringe of artificial intelligence, cognitive science and Big Data, cognitive networks are quantitative, large-scale representations of associative knowledge in the human mind, in the so-called *mental lexicon*, a cognitive system storing and processing ideas through conceptual associations and features expressible in language (Vitevitch 2019, Siew et al. 2019, Castro and Siew 2020, Vivas et al. 2020).

This perspective article reviews the most relevant recent results on cognitive networks and links them to social media in order to obtain novel insights on cognition. Cognitive networks provide quantitative readings of people's minds through their language, thus leading to next-generation algorithms and interpretable models capable of using social media data for grasping mechanisms like attention, perception and memory (Hills et al. 2019, Stella 2020b, Radicioni et al. 2020). Whereas machine learning is powerful in tagging stances or sentiment patterns (Mohammad et al. 2016), cognitive networks offer a higher interpretability, providing direct and transparent access to the structure of knowledge as embedded in social media content like news, tweets or posts.

This article will identify key potential developments of cognitive network science in relation with the above five cognitive dimensions of social media. Particular attention will be devoted also to the possibility of using multilayer networks for encapsulating within the same framework both social ties and cognitive relationships. Data limitations and ethical monitoring will be discussed in view of relevant research.

**Cognitive Networks as Models of the Human Mind**

The knowledge humans use for producing social media messages is mostly linguistic and resides in the so-called mental lexicon, a cognitive system apt at acquiring, storing, processing and producing conceptual knowledge (Vitevitch 2019). Despite its name, it is no "common" dictionary but rather represents a highly dynamical and structured system of conceptually interconnected ideas, whose access and testing cannot thus be mediated in a lab. Whereas experimenters could physically manipulate a human brain in the lab in order to test its connections, the mental lexicon remains a cognitive construct, with fascinating influence over the mechanisms of information processing (cf. Vitevitch 2019, Castro and Siew 2020). For this reason, the structure of knowledge in the human mind has to be investigated in other ways, like through cognitive tasks stimulating the mental lexicon (e.g. people writing about a topic) or through representations of conceptual knowledge in the mental lexicon. Cognitive networks can combine both these types of indirect access, as they can be built either through cognitive tasks or as representations of specific semantic, orthographic, phonological, syntactic or even visual aspects of knowledge in the mental lexicon (Siew et al. 2019).

Conceptual networks were first proposed as models of knowledge in the human mind by Quillian (1967), in a hierarchical organisation of concepts interconnected when sharing semantic features. Network distance on such a structure could account for the time it took for participants to rate the validity of simple statements but not other patterns related to meaning negation. Given the scarcity of datasets at that time, cognitive networks were rapidly forgotten (Castro and Siew, 2020). Recently, cognitive networks were rediscovered, thanks to the advent of Big Data and novel theoretical tools from network scientists and physicists. Nowadays, complex networks capturing hundreds of thousands of semantic, phonological, orthographic and syntactic similarities are increasingly successful in both descriptive and predictive models of language acquisition (Hills et al. 2009, Stella et al. 2017), lexical processing (Kenett et al. 2017, Vitevitch 2019, Murphy et al. 2020, Vivas et al. 2020) and cognitive degradation (Castro et al. 2020).

**Information on social media: Language is key but multifaceted**

Reading sentences, much like this one, activates the mental lexicon and its cognitive structure of interconnected meanings and emotions (Vitevitch 2019). Analogously, the mental lexicon is used by social media users whenever they express their knowledge through online messages. In this way, language represents a powerful bridge between the online world, the way humans express themselves and the way their minds are organised.

A first issue for investigating social discourse is represented by the massive amounts of available data (Krippendorf, 2018). For instance, only on Twitter, online users produce over 6,000 tweets in one second (cf. Brandwatch.com). This deluge of information contains various types of content: written language, hyperlinks, emojis, pictures and videos (see also Figure 1). Even if one discarded pictures and videos, whose automatic processing remains an open challenge, a rich linguistic information would remain in the form of words, hashtags, social jargon and emojis. All these elements provide key semantic and emotional cues used by social users to express themselves. In particular, even though hashtags and emojis are not a language, e.g. they do not satisfy grammatical rules, they can still contain relevant info. This multifaceted nature of online social media messages can be accounted for in cognitive networks, where words, hashtags, jargon and emojis can be represented as network nodes bringing potentially different types of information. Hashtags can provide information about the general topic of messages (Gallagher et al. 2018, Mehler et al. 2020) and thus enable a selection of messages around specific topics. Alternatively, hashtags and emojis can also be translated in words, enriching the overall semantic content of posts (cf. Stella et al. 2020).

**Cognitive networks, social media data and the need for interpretability**

To achieve next-generation tools suitable for processing and interpreting knowledge, it is fundamental to combine together automatic tools and cognitive models (Krippendorf et al. 2018, Mehler et al. 2020). On the one hand, cognitive science can rely on psychology and neuroscience in order to identify theoretical aspects of information processing, however it cannot test its theories without access to large-scale, longitudinal and context-dependent datasets (Li et al. 2019, Castro and Siew 2020). On the other hand, AI and machine learning are proficient at spotting knowledge patterns and correlations often invisible to the human eye in large datasets (Mohammad et al. 2016), though the clear mechanisms behind such performance often lack interpretability and clear-cut cognitive grounding (Rudin et al. 2019). For instance, identifying positive/negative stances could be solved with high accuracy by machine learning but as a “black-box”, i.e. with no insights about the cognitive structure characterising stances themselves (Rudin et al. 2019).

Cognitive network science lies in between cognitive models and Big Data and it shows how representing knowledge via conceptual associations can offer interpretable insights. An example is semantic network distance, i.e. the smallest number of conceptual associations linking two concepts. Recently, Kenett and colleagues (2017) showed that semantic network distance outmatched latent semantic analysis in predicting human judgements about semantic relatedness, with the advantage of semantic distance being interpretable through spreading activation (Quillian, 1967).

**Cognitive networks and the “text as data” approach**

Humans do not explicitly see conceptual associations when reading sentences and yet they are aware of the syntactic and semantic links combining information units, e.g. words, and conferring meaning and emotions to a given sentence (Krippendorf 2018, Vivas et al. 2020). These links, present in the structure of knowledge, are mirrored in the networked structure of the mental lexicon, so that language is informative of the cognitive reflection and structured mindsets of individuals (Vitevitch 2019). Several ways for giving structure to knowledge through cognitive networks have been introduced in the past.

Word co-occurrences have been one of the first approaches detecting knowledge structure from language (cf. i Cancho et al. 2001). Words co-occur when appearing one after each other or within a sequence of *L* of words. Co-occurrences are simple to compute but come at the cost of considering noise in the form of relationships with language units without intrinsic meaning (i.e. stopwords, like prepositions). Approaches relying on word co-occurrences but carefully managing (Quispe et al. 2020) or removing stopwords (Stella et al. 2020) have been successful in several tasks modelling the structure of knowledge in texts. Co-occurrence networks detected key concepts in texts (Matsuo and Ishizuka, 2004), identified key aspects of social protest about COVID-19 and national lockdown in online social discourse (cf. Stella et al. 2020, Stella 2020b) and predicted company stock prices through discussion in online forums (Colladon and Scettri 2019). Stella et al. (2018) used co-occurrences of hashtags for exposing how accounts driven by automatic software injected hate speech in online discourse around the Catalan referendum in 2017. Amancio and colleagues (2012) showed that co-occurrence networks highlighted dissimilarities between authors’ writing styles, thus facilitating their identification. Considering multiplex networks of co-occurrences and feature sharing improved author identification accuracy (Quispe et al. 2020). These results indicate already simple-to-parse word co-occurrences can highlight relevant structural features of posts and texts.

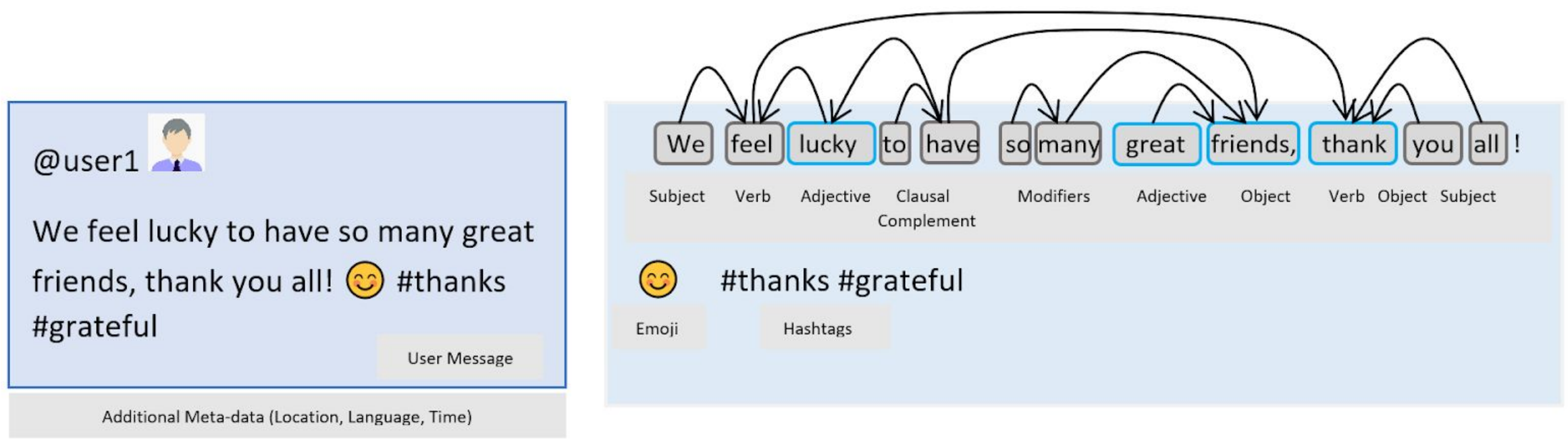

**Figure 1: When reading sentences, we do not see connections between concepts and yet we activate these links to process meaning. Analogously, a user's post/tweet/message (left) contains syntactic, semantic and emotional networked information (right). Language on social media can also include emojis and hashtags. Positive (neutral) words were highlighted in blue (grey). Syntactic dependencies and part-of-speech tagging can be reconstructed via machine learning (here TextStructure[] by WolframResearch was used) and form a complex network analogous to textual forma mentis networks (cf. Stella 2020).**

The need to focus specifically on the syntactic structure of language and the availability of increasingly more accurate AI models for extractic syntactic dependencies both motivated the development of syntactic cognitive networks (cf. i Cancho et al. 2004). Syntactic relationships focus over combinations of words that provide, together, a given meaning to a certain sentence or piece of text (see also Figure 1). Extensive research has shown that the layout of syntactic links between words is indicative of a dependency distance minimisation, i.e. a cognitive mechanism at work in the mental lexicon and actively placing syntactically related words close to each other in sentences (Ferrer-i-Cancho and Gomez-Rodriguez 2019). These results further highlight the cognitive relationship between the structure of syntactic relationships and cognitive patterns. Building on this cognitive interpretation, Stella (2020) enriched syntactic networks with synonym relationships in order to produce textual forma mentis networks (TFMNs), reconstructing the syntactic and semantic knowledge in texts reminiscent of authors' mental lexicon organisation. Stella showed that TFMNs were capable of detecting key concepts in both annotated short texts and social media data, providing ways of capturing semantic prominence in texts without word frequency.

The above results open promising directions for further using networks of concepts when investigating social cognitions in online platforms, which is briefly reviewed in the following in terms of exploring cognitive phenomena related to salience, perception and biases.

**Studying semantic salience beyond frequency counts**

Salience is the state of being prominent in a given context. Semantic salience or prominence characterises concepts that are key for individuals to understand a greater extent of knowledge and its influence over cognitive representations is a crucial research direction (cf. Vivas et al. 2020).

Social discourse can provide data useful for understanding those features influencing semantic prominence and ultimately determining key aspects of social discourse.

A key feature for identifying semantic prominence of individual concepts through social media is frequency, i.e. counting how many times a given word occurs. This metric influences a variety of linguistic tasks (cf. Vitevitch 2019) but it is based on concepts in isolation. A word might occur more or less frequently but also always within the same context or not. Frequency alone would not be able to capture these differences. For this reason, n-grams were introduced in the literature as frequency counts of words together with other n-1 contextually related words (Damashek, 1995). n-grams can account for contextual information but also require prior knowledge on how to select words.

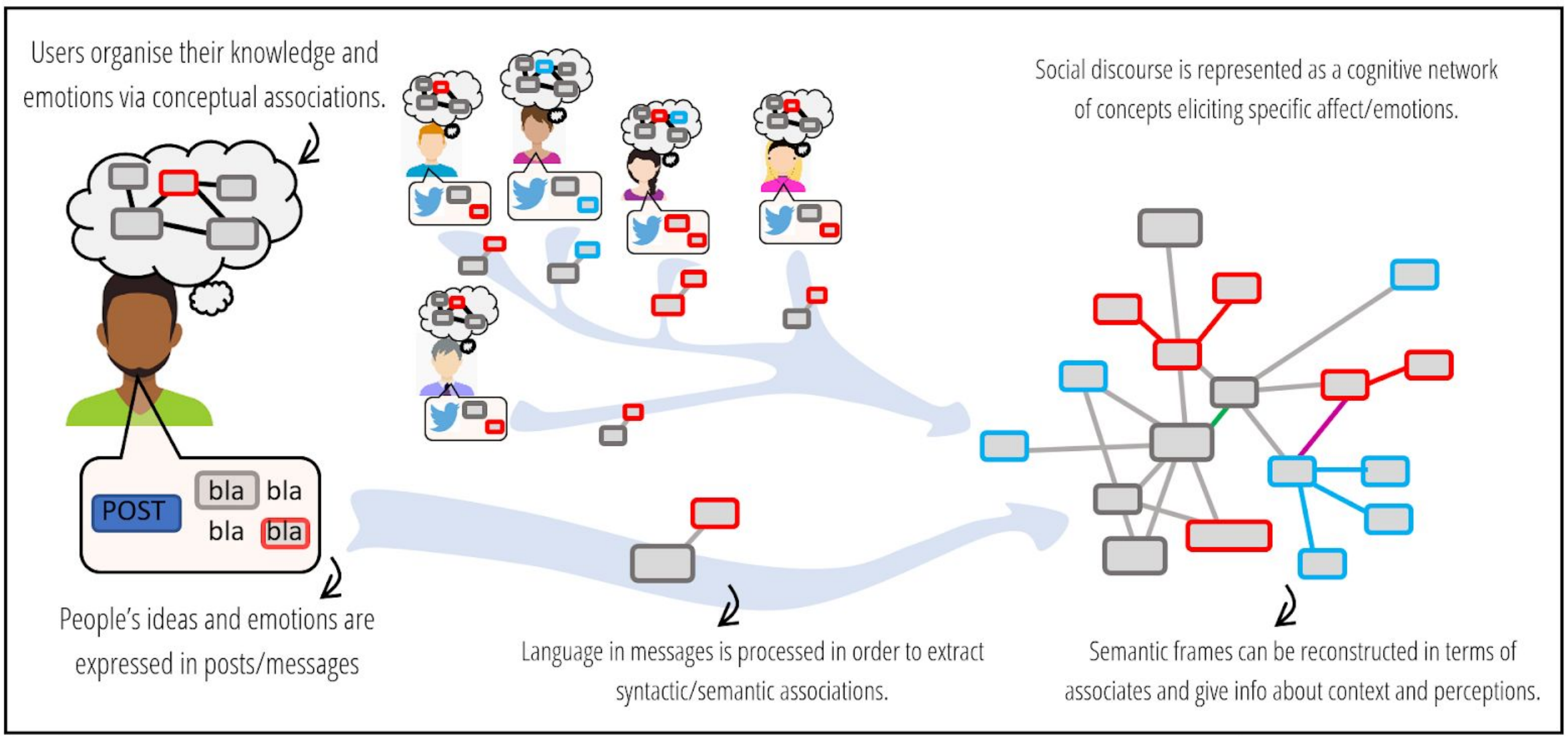

**Figure 2: Infographics about building cognitive networks out of users' messages in social media. The final cognitive network can give structure to the ideas and emotions expressed by users in social discourse as a network with links of different types (e.g. multilayer structure) and/or words expressing different emotions/sentiment (represented here by coloured frames).**

Keeping *n* fixed to 2 but considering pairwise associations among more words leads to co-occurrence networks. Further filtering out non-syntactic links leads to syntactic networks. In this way, syntactic networks are an extension of mere frequency counts (1-grams), considering how conceptual relationships are structured across different contexts or semantic areas, i.e. clusters of words with analogous semantic features (Citraro et al. 2020).

Representing social discourse as a syntactic/semantic network can be a quantitative way of estimating concept salience/prominence in social media data beyond frequency (see also Figure 2). Prominence would be defined in terms of the semantic relatedness of a concept across contexts, with more prominent concepts being closer to and more well connected than others. Stella (2020b) built on previous results of semantic relatedness/closeness being captured by semantic network distance (Kenett et al. 2017, Kenett et al. 2018) in order to model semantic prominence in terms of closeness centrality, i.e. the mean network distance of one node to all its connected neighbours. The author found that frequency counts and closeness identified different classes of concepts in social discourse about COVID-19. Negative jargon, expressing social protest and complaint, was more frequent but appeared with a lower closeness centrality, i.e. it was more peripheral and less well connected across the different semantic contexts debated by users. Instead, hopeful jargon about the reopening appeared with less frequency but with a higher closeness centrality, indicating a higher structural complexity and richness of different contexts surrounding this jargon in social discourse. Importantly, word frequency was not capable of detecting fluctuations in semantic prominence that were evident with closeness centrality and which reflected news media announcements about the quarantine in Italy (cf. Stella 2020b).

Semantic salience/prominence can be investigated also by using multiple network metrics at once, like in the study by Gallagher and colleagues (2018) about the #BlackLivesMatter and #AllLivesMatter movements. Through a backbone filtering of spurious co-occurrences and local and global network metrics, the authors showed that social discourse around #BlackLivesMatter contained a richer number of semantically unrelated but salient concepts in comparison to the language in messages with #AllLivesMatter. #BlackLivesMatter displayed more diverse, multifaceted and semantically richer discourse structure than #AllLivesMatter.

Identifying prominent concepts of social discourse is key for understanding the narratives promoted across large audiences online (Gallagher et al. 2018, Stella et al. 2020) and can provide datasets useful for future investigations studying the cognitive mechanisms characterising how semantic prominence in such concepts originated. The above results call for future research in this direction, going beyond frequency measures and including accessible contextual information.

**Give structure to online semantic frames and emotional profiles**

If the network structure can be used for achieving models explaining semantic salience in terms of large-scale conceptual associations, can networks and social media data assist also with the study of perception? A quantitative model for studying perception through social media data should be related to understanding how social media users perceive events through their online discussions. Cognitive networks provide a convenient and powerful way for reconstructing quantitatively the semantic patterns framing ideas on social media discours thanks to the theory of frame semantics (cf. Fillmore, 2001). As also highlighted in Figure 1, meaning is cast upon individual words by conceptual associates referring to them. Semantic frame theory posits that the idea and features of a given entity (say "pandemic") can be reconstructed by considering the other words referring to it syntactically and semantically in a given text (e.g. "people", "infection", "hospitals", "populations", "large") (Baker et al. 1998, Fillmore 2001). Semantic frames thus surround a given word, including its related concepts, e.g. people, places and situations. In resources like FrameNet (cf. Baker et al. 1998), semantic frames are complete only when domain knowledge is made explicit, which is not always the case in social media data.

Cognitive networks can identify semantic frames as communities of tightly interconnected concepts (Correa and Amancio, 2019; Ribeiro et al. 2020; Citraro and Rossetti 2020) or network neighbourhoods (Stella 2020), the latter being extensively used in psycholinguistic inquiries about language processing (cf. Vitevitch 2019). In this way, the semantic frame of a given word *W* can be considered either as the network neighbourhood of syntactic/semantic associates of *W* or the network community of other words tightly connected with *W* and between each other. Figure 2 reports how different semantic frames can be obtained from social media discourse.

Investigating perception could then be recast into the problem of analysing network neighbourhoods of semantic/syntactic associates around specific topics of social discourse. Pioneering results towards this direction already show that semantic frames can contain key information about the way social users discuss and perceive their experience and beliefs. Stella (2020) found that online discourse framed the idea of the "gender gap", usually highly criticised, in a context rich with positive jargon, evoking emotions of joy and trust and celebrating women's success in science. Figure 3 reports the semantic frame (left) and emotional profile (right) of "bias" as reported by (Stella 2020). Online users expressed awareness about the need to fight unconscious biases through jargon eliciting more trust and anger than expected at random. An advantage of representing semantic frames as networks is the immediate visualisation of the content and emotions surrounding specific aspects of discourse.

Semantic frames could be reconstructed also in terms of network communities, as suggested by recent work including linguistic features in community detection analysis and identifying coherent semantic frames from cognitive network data (Citraro and Rossetti, 2020). Radicioni and colleagues (2020) identified semantic frames as communities in hashtag co-occurrence networks. The authors reconstructed how users with different political views semantically framed the Italian elections in 2018. Media coverage was found to strongly polarise frames, enriching them with jargon either strongly supporting or deeply criticising political figures. Such influence might be explained by the great deal of public attention directed towards online media coverage, as recently found in (Perra et al. 2020). The above pioneering results indicate that the investigation of semantic frames on social media can capture the cognitive influence exerted by news media over polarising social users or altering their perceptions, a crucial yet relatively unexplored research field.

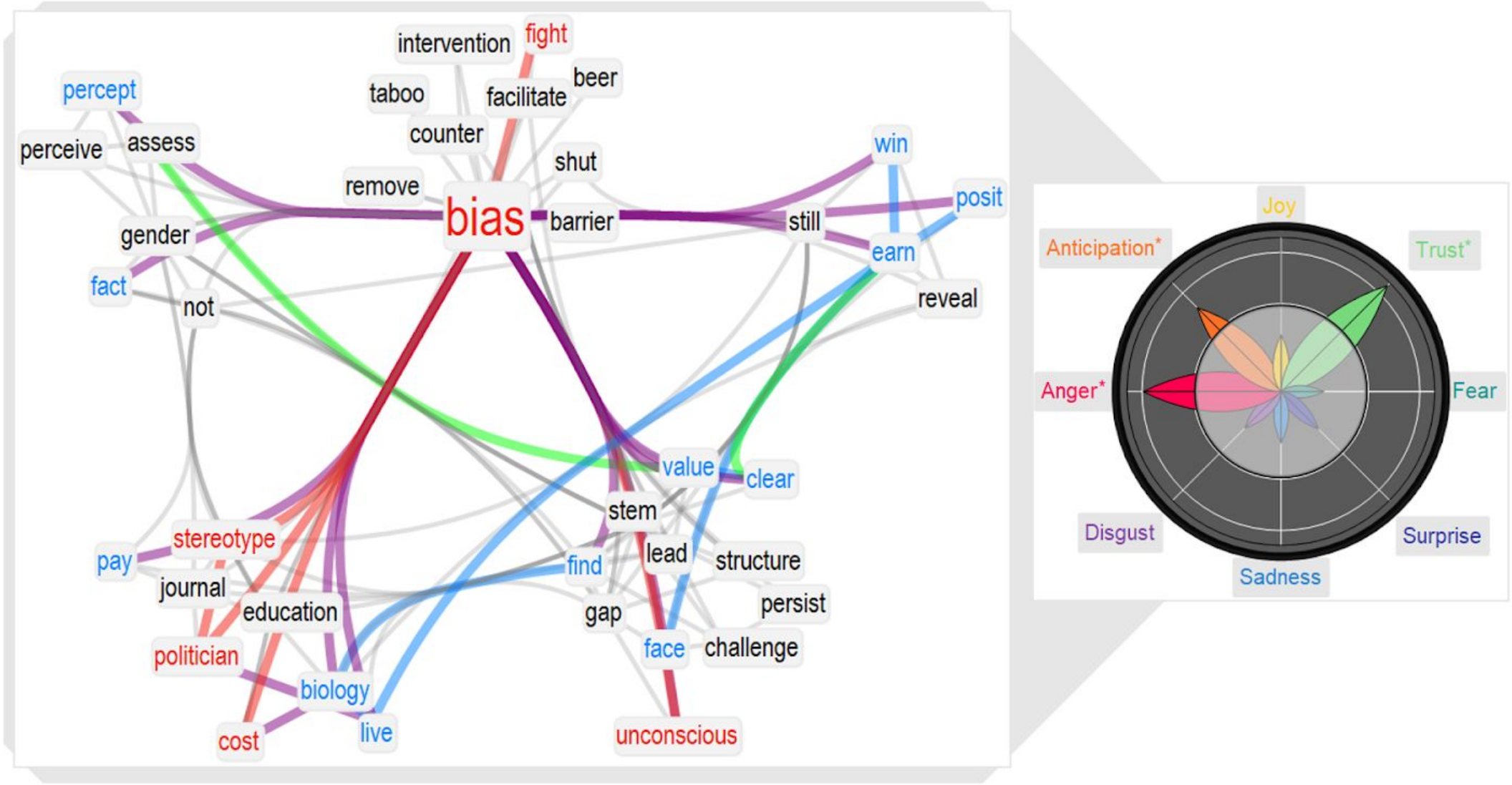

**Figure 3: Right: From social discourse about the gender gap, semantic frame of syntactic (red, cyan and grey) and semantic (green) conceptual associations around "bias". Semantic areas are represented as word clusters, identified through Louvain community detection. Left: Emotional flower, reporting the z-score of the observed number of words eliciting a certain emotion in comparison to random expectation. Petals falling outside the white circle indicate z-scores higher than 1.96, i.e. emotions stronger than random expectations (cf. Stella 2020).**

Reconstructing semantic frames from syntactic/semantic dependencies in text is an important task also for unearthing how concepts are structured and promoted by fake news and misinformation channels in the "dark side" of information flow (Hills 2019). In their investigation of conspiratorial theories and the 5G on Twitter, Ahmed and colleagues (2020) found that websites promoting fake news were largely reshared by users framing negative stances towards conspiratorial theories. Despite criticising fake content, these users ultimately promoted fake news through online discussion. Combined with other recent findings that reliable and questionable sources of news can spread in different ways on Twitter (Pierri et al. 2020) but not on other social media platforms (Cinelli et al. 2020), then detecting online controversial content calls for additional consideration of the cognitive dimension of users' semantic frames within future research. Attention should be devoted to going beyond resharing counts, understanding the ways information is cognitively perceived, framed and subsequently diffused.

Investigating semantic frames is important also for monitoring aims, like identifying the overall wellbeing and mental health of large audiences (Linthicum et al. 2019). Cognitive network science has already been used in investigations related to mental health. Forbush et al. (2016) used cognitive networks for reconstructing the symptoms expressed by people with eating disorders. The authors gave structure to the correlations between symptoms as expressed by 143 individuals and found "body checking" as the most semantically prominent, i.e. a key symptom to act upon for enhanced cognitive-behavioural therapies. Analogously, a recent textual analysis of 139 suicide notes showed that "love" and "life" are key to those who commit suicide, although the latter framed such words with drastically different sets of emotions when compared to mindwandering in absence of suicide ideation (Teixeira et al. 2020). These network-powered results are promising in identifying novel features of texts and posts for better identification of mental health issues in social media posts (see also Linthicum et al. 2019).

**Profiling online users by considering the language they used in posts**

The cognitive dimension of messages is key not only for investigating people's perceptions but also for studying salient features of users' cognitions like attitudes or personality traits, which all contribute to the creation of a "cognitive fingerprint" identifying how individuals reason and behave through associative networks (cf. Abbas et al. 2020). Cognitive network science has shown that giving structure to knowledge can uncover patterns of relevance for personality traits like creativity (Mednick 1962, Kenett et al. 2014), curiosity (Lydon-Staley et al. 2020) and openness-to-experience (Christensen et al. 2018).

Generally speaking, creativity is the ability to draw connections among apparently unrelated concepts, filling gaps in knowledge or finding new ones (Mednick 1962). Several recent works have shown that the structure of knowledge in the mental lexicon is highly predictive of the levels of creativity in populations (Kenett et al. 2014, Benedek et al. 2017, Valba et al. 2020) and individuals (Kenett et al. 2018). For instance, Kenett and colleagues (2018) showed that people with a higher creativity level were better at connecting concepts from different semantic communities, thus enhancing the connectivity of their mental lexicon even under progressive word removal. Always using semantic networks, Valba and colleagues (2020) found that associating more remote concepts activated mainly weaker conceptual links, connecting different semantic communities. These studies indicate that people with a higher creativity can navigate their knowledge in different ways compared to people with lower creativity and this difference can be found in language. Reconstructing semantic/syntactic networks in social media, potentially through textual forma mentis networks (Stella, 2020), would therefore open concrete, quantitative ways for novel research directions where the creativity of online users is estimated by simply considering their messages, calling for future research endeavours.

In a similar fashion, also other personality traits might be estimated from users' language, like openness to experience (Christensen et al. 2018, Christensen et al. 2019) and curiosity (Lydon-Staley et al. 2020). Openness to experience identifies people's willingness to embrace change, be intellectually curious and open-minded. This definition was provided by Christensen and colleagues (2019) through a network community analysis over different taxonomies used for measuring openness itself. Like for curiosity (Kenett et al. 2014), individuals with a higher openness to experience exhibited a different organisation of associative knowledge in their mental lexica in comparison to people with lower openness (Christensen et al. 2018), thus opening new research opportunities for studying openness to experience in online contexts thanks to social media data and cognitive network science.

A different approach could be used for determining individuals' curiosity, a tendency to seek information due to multi-faceted feelings of deprivation (Lydon-Staley et al. 2020). The recent study by Lydon-Staley and colleagues (2020) built semantic networks out of Wikipedia webpages visited by individuals who were also monitored through self-assessed surveys. Participants with a strong drive to know and eliminate gaps in their knowledge ended up building considerably more tightly interconnected semantic networks than other individuals with a lower sense of curiosity. Also different behavioural patterns of information seeking and network navigation were found. Individuals tended to either revisit concepts many times, intensifying local conceptual associations ("hunters"), or to explore knowledge by jumping across different semantic areas ("busybodies"). Understanding how people revisit concepts in subsequent messages (e.g. a Twitter timeline) in a way similar to (Lydon-Staley et al. 2020) might provide additional cognitive footprints about curiosity and information seeking mechanisms. An analogous procedure, considering only conceptual revisiting of the same concepts and semantic areas in tweets was introduced by Monakhov (2020) for detecting trolls, i.e. users inflaming social discourse with abrasive language and influencing even massive voting events or public perceptions (Broniatowski et al. 2018). Monakhov (2020) showed that trolls tend to behave like

Lydon-Staley and colleagues' "hunters", revisiting only a limited span of conceptual associations, a feature that was used for detecting trolling with an accuracy of 91% but using only 50 tweets per user.

**Multilayer networks as socio-cognitive models**

Most of the above works focused only on one type of conceptual associations despite the mental lexicon accounting for multiple types or layers of conceptual relationships, e.g. syntactic, semantic, visual and many others. Multilayer networks (cf. Boccaletti et al. 2014) can account for multiple types of links within one network representation. Besides textual forma mentis networks (Stella 2020, Stella 2020b), as described above, other cognitive approaches with multilayer networks were recently introduced in the literature. Conceptual distance in multilayer networks (combining phonological, categorical and semantic associations) was found to be predictive of both normative word acquisition in children (Stella et al. 2017) and picture naming failures in people with a spectrum of aphasic disorders (Castro et al. 2020). Combining phonological and orthographic similarities in a phonographic multilayer network, Siew and Vitevitch showed that network degree predicts a facilitatory effect for visual word recognition absent in single-layer networks (Siew and Vitevitch 2019). By connecting topics according to semantic/syntactic similarity, Mehler and colleagues introduced *multiplex topic networks*, which highlighted a stronger co-occurrence of geographically distant places featuring analogous semantic features in online social discourse (Mehler et al. 2020).

For their ability to capture different types of links among the same or potentially different groups of nodes, multilayer networks possess the potential to model both social and cognitive interactions among users. An example is Pietrocola and Rodrigues's work (Pietrocola and Rodrigues 2020), which combined networks of conceptual associations with networks of social interactions. In a classroom setting, the authors showed that students connected by stronger social ties also exhibited similarly structured conceptual networks around the topics of classroom life. Such correspondence between similar knowledge and stronger social ties in social media relates to echo chambers, i.e. groups of actively interacting online users with the same stance (Del Vicario et al. 2016, Cinelli et al. 2020). In turn, echo chambers are caused by confirmation bias, i.e. a cognitive tendency for the human mind to simplify the acquisition of low-quality new knowledge by adjusting it to a pre-existing set of beliefs and info (Hills 2019). Overloaded by brief and often incomplete information (Qiu et al. 2017), online users end up being more susceptible to confirmation bias and thus acquire mostly information already suiting their pre-existing knowledge (Hills 2019). Multilayer networks would open novel quantitative scenarios for the investigation of social cognition within such echo chambers.

## DISCUSSION AND OPEN CHALLENGES

Beyond the methodological gaps and research opportunities outlined above, the framework of cognitive network science for investigating cognition through social media poses also important meta-research advantages and limitations.

The main advantage of cognitive networks is their intrinsic ability to provide jargon and quantitative metrics that can be interpreted from a cognitive perspective (Siew et al. 2019). For instance, "semantic relatedness" can be translated, quantified and explained in terms of network distance (Kenett et al. 2017) and this, in turn, gives rise to distance-based measures that can complement frequency and n-grams in detecting trending online topics (Stella 2020b). Bridging Big Data analytics with interpretable modelling through cognitive networks can greatly benefit several research lines studying cognition like semantic framing, stance detection and perception, emotional dynamics and cognitive biases, as outlined above.

Cognitive networks can be powerful assets also for exploring the cognitions portrayed by AI-driven accounts like social bots (Ferrara et al. 2016). Often posing as human accounts, social bots are driven by automatic code and promote human-like content, so that the above network techniques would work also for better understanding what social bots do. By considering cognitive networks of hashtag co-occurrences and networks of social interactions, combined, (Stella et al. 2018) found that social bots mainly targeted influential human users and introduced hate speech in specific online groups. Characterising the structure of knowledge promoted by social bots through cognitive networks represents a key way for unveiling how automated accounts can manipulate human users, their perceptions and their information seeking behaviour (Ferrara et al. 2016, Broniatowski et al. 2018, Hills 2019).

Cognitive networks can also enable novel ways of profiling users according to their language, like discussed above. These pioneering approaches are promisingly powerful but pose important limitations in terms of representativeness, variability and integration. Social media can give voice to large audiences but not to everyone, as it is increasingly clear that only certain demographic groups access online platforms, e.g. older people with a lower income tend to tweet less than younger professionals (Brandwatch.com). This poses important limitations to the extent to which social media inquiries provide access to data sufficient enough for fully characterising both global perceptions and individual users'. Related to this issue there is individual variability. Although sometimes users achieve consensus and provide a normative, average perception representative of the whole group (Cinelli et al. 2020), this is not always the case. Even at the individual level, a person might display flickering emotions (Stella 2020b), change their language over time (Dodds et al. 2011) or across different platforms/social contexts (Cinelli et al. 2020). This underlines the importance of integrating data from multiple sources, providing monitoring tools that sample across different platforms and over time in order to read individual users' mindsets in their online social environments. Preliminary results with trolling (Monakhov 2020) indicate that little individual data is required for profiling some users' behaviour but additional research about personality traits and social media is required.

There are several ethical questions. Users adopt the online world often as a way to communicate and find meaning in their own personal lives, a phenomenon that has been studied for decades by narrative psychology (Laszlo 2008). If online users exploit their own and others' narratives in order to deal with experiences, gathering their online data even only for research purposes might represent a violation of a delicate personal space. The possibility of accessing someone's mind and personality through their social data poses novel risks for privacy. Would it be ethical for a research institution to gauge online audiences and perform business predictions about, say, the job market without explicitly asking for consent to the users themselves? Considering confirmation bias, information pollution produced by social bots and misinformation campaigns (Hills 2019), would the data processed by online ecosystems be even suitable for commercial purposes? Responsible research requires important ethical guidelines covering the ever-increasing opportunities of social media investigations provided by cognitive network science and outlined here. Reading people's minds through networks calls for new research opportunities but also for additional ethics efforts. Cognitive-, data-, social- and network scientists should contribute to the quantitative development of cognitive networks for studying cognition through social media data. However, these scientific figures should also work together with philosophers and law experts under the umbrella of complexity science in order to provide clearer indications for an ethical growth of this whole new field, taking inspiration from recent efforts in organisational psychology and network science (Cronin et al. 2020).

Social media data and cognitive networks can revolutionise current models and understanding of cognition across the areas concretely outlined in here. This opens to a brave new world of opportunities for a new generation of researchers bridging different fields and pioneering whole new ways of reading, understanding and even predicting people's minds through digital media.